\documentstyle[12pt]{article}
\newcommand\bea{\begin{eqnarray}}
\newcommand\eea{\end{eqnarray}}
\newcommand\g{\gamma}
\newcommand\s{\sigma}
\begin{document}
\thispagestyle{empty}
\bibliographystyle{unsrt}
\setlength{\baselineskip}{18pt}
\parindent 24pt
\vspace{60pt}


\begin{center}{
{\Large {{\bf Purity and decoherence\\
in the theory of damped harmonic oscillator }}
}
\vskip 1truecm
A. Isar${\dagger\ddagger}^{(a)}$, A. Sandulescu${\dagger\ddagger}$
and W. Scheid${\ddagger}$\\
$\dagger${\it Department of Theoretical Physics, Institute of
Atomic Physics \\
Bucharest-Magurele, Romania }\\
$\ddagger${\it Institut f\"ur Theoretische Physik der
Justus-Liebig-Universit\"at \\
Giessen, Germany }\\
}
\end{center}

\vskip 1truecm
\begin{abstract}
For the generalized master equations derived by Karrlein and
Grabert for the microscopic model of a damped harmonic oscillator,
the conditions for purity of states are written, in particular for
different initial conditions and different types of damping,
including Ohmic, Drude and weak coupling cases, Agarwal and
Weidlich-Haake models. It is shown that the states which remain
pure are the squeezed states with constant in time variances. For
pure states, the generalized nonlinear Schr\" odinger-type
equations corresponding to these master equations are also
obtained. Then the condition for purity of states of a damped
harmonic oscillator is considered in the framework of Lindblad
theory for open quantum systems. For a special choice of the
environment coefficients, the correlated coherent states with
constant variances and covariance are shown to be the only states
which remain pure all the time during the evolution of the
considered system. In Karrlein-Grabert and Lindblad models, as
well as in the considered particular models, the expressions of
the rate of entropy production is written and it is shown that the
states which preserve their purity in time are also the states
which minimize the entropy production and, therefore, they are the
most stable ones under evolution in the presence of the
environment and play an important role in the description of
decoherence phenomenon.
\end{abstract}
\vskip 0.5truecm



\vskip 0.5truecm (a) e-mail address: isar@theor1.theory.nipne.ro



\section{Introduction}

In the last two decades, more and more interest has arisen about the search
for a consistent description of open quantum systems [1--5]
(for a recent review
see Ref. \cite{rev}).
Dissipation
in an open system results from microscopic reversible interactions between
the observable system and the environment.
Because dissipative processes imply irreversibility and,
therefore, a preferred direction in time, it is generally thought that quantum
dynamical semigroups are the basic tools to introduce dissipation in quantum
mechanics. In Markov approximation and for weakly damped systems,
the most general form of the generators of such semigroups was given
by Lindblad \cite{l1}. This formalism has been studied for the case of damped
harmonic oscillators [6, 8--12]
and applied to various physical phenomena, for instance, to the
damping of collective modes in deep inelastic collisions in
nuclear physics \cite{i1}. A phase space representation for open
quantum systems within the Lindblad theory was given in
\cite{i2,vlas}. Important progress beyond the limitations of the
weak coupling approach was made in describing quantum dissipative
systems by using path integral techniques \cite{weis,karr}.

In the present study we are also concerned with the observable
system of a harmonic oscillator which interacts with the
environment. We discuss under what conditions the open system can
be described by a quantum mechanical pure state. In Sec. 2 we
present the generalized uncertainty relations and the correlated
coherent states, first introduced in \cite{dodkur}, which minimize
these relations. In Sec. 3 we consider generalized quantum master
equations derived by Karrlein and Grabert in Ref. \cite{karr} for
the microscopic model of a harmonic oscillator coupled to a
harmonic bath \cite{grab} by using the path integral and we obtain
conditions for the purity of states, in particular for different
initial conditions and different types of damping, including
Ohmic, Drude and weak coupling cases, Agarwal and Weidlich-Haake
models. We show that the states which satisfy the conditions of
purity are the pure squeezed states with well-determined constant
in time variances. For pure states, we also derive the generalized
Schr\" odinger-type nonlinear equations corresponding to these
master equations. The Lindblad theory for open quantum systems is
considered in Sec. 4. For the one-dimensional harmonic oscillator
as an open system, we show that for a special choice of the
diffusion coefficients, the correlated coherent states, taken as
initial states, remain pure for all time during the evolution. In
some other simple models of the damped harmonic oscillator in the
framework of quantum statistical theory \cite{zel,hug}, it was
shown that the Glauber coherent states remain pure during the
evolution and in all other cases the oscillator immediately
evolves into mixtures. In this respect we generalize this result
and also our previous result from \cite{ass} as well as the
results of other authors \cite{halzou}, obtained by using
different methods. In Sec. 5 we introduce the linear entropy, we
present its role for the description of the decoherence phenomenon
and we also derive the expressions of the rate of entropy
production. We show in Lindblad and Karrlein-Grabert models that
the correlated coherent states, respectively the pure squeezed
states, which fulfill the condition for purity of states, are also
the most stable states under evolution in the presence of the
environment and make the connection with the work done in this
field by other authors
[23--28]. Finally, a summary is given in Sec. 6.

\section {Generalized uncertainty relations}
In the following we denote by $\sigma_{AA}=<\hat A^2>-<\hat A>^2$
the dispersion of the operator $\hat A$, where $<\hat
A>\equiv\sigma_A={\rm Tr}(\hat\rho \hat A), {\rm Tr \hat\rho}=1$
and $\hat\rho$ is the statistical operator (density matrix). By
$\sigma_{AB}=(1/2)<\hat A\hat B+\hat B\hat A>-<\hat A><\hat B>$ we
denote the correlation (covariance) of the operators $\hat A$ and
$\hat B.$ Schr\" odinger \cite{schr} and Robertson \cite{rob}
proved that for any Hermitian operators $\hat A$ and $\hat B$ and
for pure quantum states the following generalized  uncertainty
relation holds: \bea \sigma_{AA}\sigma_{BB}-\sigma_{AB}^2\ge
{1\over 4}|<[\hat A,\hat B]>|^2. \label{rob}\eea For the
particular case of the operators of momentum $\hat p$ and
coordinate $\hat q$ the uncertainty relation (\ref{rob}) becomes
\bea \sigma\equiv\sigma_{pp}\sigma_{qq}-\sigma_{pq}^2
\ge{\hbar^2\over 4}. \label{genun}\eea This result was generalized
for arbitrary operators (in general non-Hermitian) and for the
most general case of mixed states in \cite{dodkur}. The inequality
(\ref{genun}) can also be represented in the form: \bea
\sigma_{pp}\sigma_{qq}\ge{\hbar^2\over 4(1-r^2)},\label{dod}\eea
where \bea r={\sigma_{pq}\over\sqrt{\sigma_{pp}\sigma_{qq}}}
\label{corcoe}\eea is the correlation coefficient. The equality in
the relation (\ref{genun}) is realized for a special class of pure
states, called correlated coherent states or squeezed coherent
states, which are represented by Gaussian wave packets in the
coordinate representation. These minimizing states, which
generalize the Glauber coherent states, are eigenstates of an
operator of the form: \bea \hat a_{r,\eta}={1\over
2\eta}[1-{ir\over (1-r^2)^{1/2}}]\hat q+ i{\eta\over\hbar}\hat p
\label{eigv}\eea with real parameters $r$ and $\eta,$ $|r|<1,
\eta=\sqrt{\sigma_{qq}}.$ Their normalized eigenfunctions, denoted
as correlated coherent states, have the form: \bea \Psi(q)={1\over
(2\pi\eta^2)^{1/4}}\exp\{-{q^2\over 4\eta^2}[1- {ir\over
(1-r^2)^{1/2}}]+{\alpha q\over\eta}-{1\over 2}(\alpha^2+
|\alpha|^2)\}, \label{eigf} \eea with $\alpha$ a complex number.
If we set $r=0$ and $\eta=(\hbar/2m\omega)^{1/2},$ where $m$ and
$\omega$ are the mass and, respectively, the frequency of the
harmonic oscillator, the states (\ref{eigf}) become the usual
Glauber coherent states. In Wigner representation, the states
(\ref{eigf}) look: \bea W_{r,\eta}(p,q)=
{1\over\pi\hbar}\exp[-{(q-\sigma_q)^2\over
2\eta^2(1-r^2)}-{2\eta^2\over\hbar^2} (p-\sigma_p)^2+{2r\over
\hbar(1-r^2)^ {1/2}}(q-\sigma_q)(p-\sigma_p)], \label{corwig}\eea
where $\s_q$ and $\s_p$ are the expectation values of coordinate
and momentum, respectively. This is the classical normal
distribution with the dispersion \bea \sigma_{qq}=\eta^2,~~
\sigma_{pp}={\hbar^2\over 4\eta^2(1-r^2)},~~ \sigma_{pq}={\hbar
r\over 2({1-r^2})^{1/2}} \label{disp} \eea and the correlation
coefficient $r.$ The Gaussian distribution (\ref{corwig}) is the
only positive Wigner distribution for a pure state \cite{huds}.
All other Wigner functions that describe pure states necessarily
take on negative values for some values of $p,q.$

In the case of relation (\ref{rob}) the equality is generally
obtained only for pure states. For any density matrix in the
coordinate representation (normalized to unity) the following
relation must be fulfilled: \bea {1\over\nu}={\rm
Tr}\hat\rho^2\leq 1. \label{ropur}\eea The quantity $\nu$
characterizes the degree of purity of the state: for pure states
$\nu=1$ and for mixed states $\nu>1.$
In the language of the Wigner function the condition (\ref{ropur})
has the form: \bea {1\over\nu}=2\pi\hbar\int W^2(p,q)dpdq\leq 1.
\label{wigpur}\eea

Let us consider the most general mixed squeezed states described by the
Wigner function of the generic Gaussian form with five real parameters:
\bea   W(p,q)={1\over 2\pi\sqrt{\sigma}}\exp\{-{1\over 2\sigma}[\sigma_{pp}
(q-\sigma_q)^2+\sigma_{qq}(p-\sigma_p)^2-2\sigma_{pq}(q-\sigma_q)(p-\sigma_p)]
\},  \label{gaus}  \eea
where $\sigma$
is the determinant of the dispersion (correlation) matrix
$\left(\matrix{\sigma_{pp}&\sigma_{pq}\cr
\sigma_{pq}&\sigma_{qq}\cr}\right). \label{sigma}$ Here, $\sigma$
is also the Wigner function area -- a measure of the phase space
area in which the Gaussian density matrix is localized.
For Gaussian states of the form (\ref{gaus}) the coefficient of
purity $\nu$ is given by \bea \nu={2\over\hbar}\sqrt{\sigma}.
\label{pursig}\eea
The inequality (\ref{genun}) must be fulfilled actually for any
states, not only Gaussian. Any Gaussian pure state minimizes the
relation (\ref{genun}). For $\sigma>\hbar^2/4$ the function
(\ref{gaus}) corresponds to mixed quantum states, while in the
case of the equality $\sigma=\hbar^2/4$ it takes the form
(\ref{corwig}) corresponding to pure correlated coherent states.

The degree of the purity of a state can also be characterized by
the quantum entropy (we put the Boltzmann's constant $k_B=1$):
\bea S=-{\rm
Tr}(\hat\rho\ln\hat\rho)=-<\ln\hat\rho>.\label{entro}\eea For
quantum pure states the entropy is identically equal to zero. It
was shown \cite{aur1,aga} that for Gaussian states with the Wigner
functions (\ref{gaus}) the entropy can be expressed through
$\sigma$ only: \bea S={\nu+1\over 2} \ln{\nu+1\over 2}
-{\nu-1\over 2}\ln{\nu-1\over
2},~~~\nu={2\over\hbar}\sqrt{\sigma}. \label{entro2}\eea

\section{Generalized quantum master equations }

In the framework of the standard microscopic model
[19, 32--34] for the damped harmonic oscillator, it was shown in
Ref. \cite{karr} that in general there exists no exact master
equation for the damped harmonic oscillator \bea
{\partial\over\partial t}\rho(t)={\cal L}\rho(t) \label{kar1}\eea
with a dissipative Liouville operator $\cal L$ describing the
dynamics of the oscillator and independent of the initial states.
For specific initial preparations the time evolution is described
exactly by a time-dependent Liouville operator and the resulting
master equation for the damped harmonic oscillator with the
Hamiltonian \bea H_{0}={1\over 2M}p^2 +{M\omega_0^2\over 2}q^2,
\label{kar2}   \eea corresponding to this Liouville operator
(given by Eq. (40) of Ref. \cite{karr}), has the following general
form ($\{,\}$ denotes the anti-commutator of two operators): \bea
\dot\rho(t)=-{i\over 2M\hbar}[p^2,\rho(t)]-{iM\over
2\hbar}\g_q(t)[q^2, \rho(t)]\nonumber\\ -{i\over
2\hbar}\g_p(t)[q,\{p,\rho(t)\}] +{M\over
\hbar^2}D_q(t)[p,[q,\rho(t)]]
-{M^2\over\hbar^2}D_p(t)[q,[q,\rho(t)]]. \label{kar3}\eea The
dependence on $\omega_0$ is included in the coefficients of
commutators.

A) For the so-called thermal initial
condition \cite{haki}, which can be used to describe initial
states of the entire system (oscillator and bath) resulting from
position measurements, $D_q(t)$ and $D_p(t)$ can be written as
\bea D_q(t)=\g_q(t)<q^2>-{<p^2>\over M^2},
~~~~D_p(t)=\g_p(t){<p^2>\over M^2}. \label{therm}\eea
Here $\g_q(t),\g_p(t)$ and the equilibrium variances $<q^2>$ and
$<p^2>$ are given in terms of the
equilibrium coordinate autocorrelation function $ <q(t)q>.$

We now derive the necessary and sufficient condition for $\rho(t)$ to be a pure
state for all times. $\rho(t)$ is a pure state if and only if
${\rm Tr}\rho^2(t)=1.$ This is equivalent with $(d/dt){\rm Tr}\rho^2(t)=0$
for all times, i. e. with the condition
${\rm Tr}(\rho(t){\cal L}\rho(t))=0.$
With the explicit form of ${\cal L}\rho(t)$ given by the
right-hand side of Eq. (\ref{kar3})
and using the
relations
$\rho^2(t)=\rho(t) \label{ro1} $ and
$ \rho(t)A\rho(t)={\rm Tr}(\rho(t)A)\rho(t),$
we obtain the following condition for a state to be pure for all times :
\bea M^2D_p(t)\sigma_{qq}(t)-MD_q(t)\sigma_{pq}(t)
-{\hbar^2\over 4}\g_p(t)=0 \label{cond1}\eea
and by inserting the expressions (\ref{therm}):
\bea \g_p(t)<p^2>\sigma_{qq}(t)-[M\g_q(t)<q^2>-{<p^2>\over M}]
\sigma_{pq}(t)-{\hbar^2\over 4}\g_p(t)=0. \label{cond2}\eea

B) For factorizing initial conditions, namely if the initial density matrix of the
entire system factorizes in
the density matrix of the oscillator and the canonical density
matrix of the unperturbed heat bath [34, 36--38],
i. e. if the oscillator and bath are uncorrelated in the initial
state, then the resulting master equation is equivalent to the
result by Haake and Reibold  \cite{haak}, who derived it directly
from microscopic dynamics and by Hu, Paz and Zhang \cite{blhu}
from the path integral representation. The condition for purity of
states for these master equations has also the form (\ref{cond2}),
where now the coefficients are given by Eqs. (87), (89) in Ref.
\cite{karr}.

For a pure state $\rho(t)=|\psi(t)><\psi(t)|,$ we can obtain from
Eq. (\ref{kar3}) the evolution equation for the wave function
$\psi(t)$ as an equation of the Schr\" odinger-type
\bea {d\psi(t)\over dt}=-{i\over\hbar}H'\psi(t).\label{schreq}\eea
Taking into account the condition for purity of states
(\ref{cond1}), we find the non-Hermitian Hamiltonian \bea
H'={p^2\over 2M}+{M\over 2}\g_q(t)q^2+{1\over
2}\g_p(t)(qp+\sigma_p(t)q- \sigma_q(t)p)\nonumber\\
+{iM\over\hbar}D_q(t)(p-\sigma_p(t))(q-\sigma_q(t))
-{iM^2\over\hbar}D_p(t)(q-\sigma_q(t))^2,\label{ham3}\eea which
depends on the wave function $\psi(t)$ via the expectation values
$\s_q$ and $\s_p,$ i. e. this Schr\" odinger-type equation is
nonlinear.

The master equations considered up to now in this Section are
exact. We now consider particular types of damping for which the
dynamics can be described in terms of approximate Liouville
evolution operators, valid for arbitrary initial states \cite{karr}.
Then the evolution operator is time-independent and the master equation
for the density matrix
obeys Eq. (\ref{kar3}), where $D_q(t)=D_q$ and $D_p(t)=D_p$
read
\bea D_q=\g_q<q^2>-{<p^2>\over M^2},
~~~~D_p=\g_p{<p^2>\over M^2}, \label{dcoef}\eea
with time-independent coefficients $\g_q$ and $\g_p$ \cite{karr}.

1) For strictly Ohmic damping,
$\g_p=\g$ is the Laplace transform of the damping kernel
\cite{karr,grab} of the model and $\g_q=\omega_0^2.$
In this case we do not have a
well-defined Liouville operator since $<p^2>$ and, therefore, the
coefficients $D_q$ and $D_p$ given by (\ref{dcoef}) are logarithmically
divergent.
The condition for purity of states for this strictly Ohmic
damping is similar to the relation (\ref{cond2}), only now the coefficients
are constant:
\bea \g<p^2>\sigma_{qq}(t)-(M\omega_0^2<q^2>-{<p^2>\over M})
\sigma_{pq}(t)-{\hbar^2\over 4}\g=0. \label{cond3}\eea

2) A more realistic case is the so-called Drude damping. For a
sufficiently large Drude parameter
 $\omega_D$ and
sufficiently high temperature $k_BT\gg\hbar\g,$ the oscillator
dynamics can be described by an approximate Liouville operator
with the coefficients \bea\g_q=\alpha^2+\eta^2,~~~~
\g_p=2\alpha,\label{drcoef}\eea where $\alpha$ and $\eta$ depend
on $\gamma, \omega_0$ and $\omega_D$
and then the condition for purity of states is
\bea 2\alpha<p^2>\sigma_{qq}(t)-[M(\alpha^2+\eta^2)<q^2>-{<p^2>\over M}]
\sigma_{pq}(t)-{\hbar^2\over 2}\alpha=0. \label{cond4}\eea
For a pure state, the Schr\" odinger equation (\ref{schreq}) corresponding to
the master equation with Ohmic damping has
the Hamiltonian (\ref{ham3}), with the coefficients given by
(\ref{dcoef}) and with
$\g_p=\g$, $\g_q=\omega_0^2$.
A similar result holds for the Drude damping, when we take the coefficients
(\ref{drcoef}).

3) In Ref. \cite{karr} it is shown
that
in the limit of weak damping the general master equation has the
following form: \bea   \dot\rho(t)= -{i\over \hbar}[{p^2\over
2M}+{M\over 2}(\omega_0^2+\omega_0\g_s)q^2,\rho(t)] \nonumber\\
-{i\g_c\over 2\hbar}[q,\{p,\rho(t)\}]- {K_s\over
M\hbar\omega_0}[p,[q,\rho(t)]] -{K_c\over\hbar}[q,[q,\rho(t)]].
\label{kar8}\eea This equation is given in terms of four
dissipation coefficients: $\g_s$ leads to a frequency shift and
may be absorbed by renormalizing $\omega_0,$ $\g_c$ is the
classical damping coefficient and the coefficients $K_s$ and $K_c$
depend on the temperature.
$K_s$ can be calculated analytically only in
certain cases. One of these is the Drude model.
The master equation (\ref{kar8}) is a generalization of the
Agarwal equation \cite{agar}: \bea   \dot\rho(t)= -{i\over
\hbar}[{p^2\over 2M}+{M\omega_0^2\over 2}q^2,\rho(t)]
-{i\kappa\over \hbar}[q,\{p,\rho(t)\}]-
\kappa{M\omega_0\over\hbar}\coth({\hbar\omega_0\over
2k_BT})[q,[q,\rho(t)]], \label{kar9}\eea which was derived with
the help of projection operator techniques from the same
microscopic model using Born approximation in conjunction with a
short memory approximation. As a main difference, in Agarwal's
equation the $K_s$ term is absent. Here $\kappa$ is a
phenomenological damping coefficient.

From Eqs. (\ref{kar8}) and (\ref{kar9}) we obtain the following
conditions for purity of states: \bea K_c\sigma_{qq}(t)+{K_s\over
M\omega_0}\sigma_{pq}(t) -{\hbar\over 4}\g_c=0 \label{cond}\eea
and, respectively, \bea M\omega_0\coth({\hbar\omega_0\over
2k_BT})\sigma_{qq}(t)= {\hbar\over 2}. \label{cond5}\eea
The corresponding Schr\" odinger-type equations for a pure state
have the Hamiltonian \bea H'={p^2\over 2M}+{M\over
2}(\omega_0^2+\omega_0\g_s)q^2+{1\over 2}
\g_c(qp+\sigma_p(t)q-\sigma_q(t)p)-iK_c(q-
\sigma_q(t))^2\nonumber\\ -{iK_s\over
M\omega_0}(p-\sigma_p(t))(q-\sigma_q(t))~~~~~~~~~~~~~~~~~~~
\label{ham4}\eea
and, respectively, \bea H'={p^2\over 2M}+{M\over 2}\omega_0^2 q^2+
\kappa(qp+\sigma_p(t)q-\sigma_q(t)p)-i\kappa
M\omega_0\coth({\hbar\omega_0 \over 2k_BT
})(q-\sigma_q(t))^2.\label{ham5}\eea

4) All the above presented time-independent Liouville operators
are not of Lindblad form.
In Ref. \cite{karr} it is shown that in the weak coupling limit,
further coarse graining will result in a Lindblad operator.
Indeed, for weak damping,
the master
equation (\ref{kar8}) simplifies and takes on the following
form, written in terms of usual creation and annihilation
operators $a^\dagger, a$:
\bea \dot\rho(t)=-i(\omega_0+{\g_s\over 2})[a^\dagger a,\rho(t)]+\g_{\uparrow}
([a^\dagger\rho(t),a]+[a^\dagger,\rho(t)a])
+\g_{\downarrow}([a\rho(t),a^\dagger]+[a,\rho(t)a^\dagger]), \label{kar10}\eea
where
\bea \g_{\downarrow,\uparrow}
={\g_c\over 4}[\coth({\hbar\omega_0\over 2k_BT})\pm 1].
\label{kar11}\eea This equation, first derived by Weidlich and
Haake \cite{weid} from a microscopic model for the damped motion
of a single mode of the electromagnetic field in a cavity, is of
Lindblad form and can be obtained formally as a particular case of
the general master equation (\ref{mast}) for the damped harmonic
oscillator (see next Section), if we take \bea D_{pp}={\hbar
M\omega_0\over 2}(\g_{\downarrow}+\g_{\uparrow}),~~
D_{qq}={\hbar\over 2M\omega_0}(\g_{\downarrow}+\g_{\uparrow}),~~
D_{pq}=0,~~\lambda=(\g_{\downarrow}-\g_{\uparrow}),~~\mu=0.
\label{kar12}\eea From Eq. (\ref{kar10}) we obtain the following
condition for purity of states:
\bea 2\coth({\hbar\omega_0\over 2k_BT})\sigma_{a^\dagger
a}(t)=1\eea or, in terms of coordinate and momentum,
\bea (M\omega_0\sigma_{qq}(t)+{\sigma_{pp}(t)\over M\omega_0})
\coth({\hbar\omega_0\over 2k_BT})=\hbar. \label{cpur} \eea For a
pure state, the Schr\" odinger-type  equation corresponding to Eq.
(\ref{kar10}) has the Hamiltonian \bea H'=H+i\hbar{\g_c\over 2}
(\sigma_{a^\dagger}a- \sigma_a a^\dagger+{1\over
2})-i\hbar{\g_c\over 2}\coth({\hbar\omega_0\over 2k_BT})
[(a^\dagger-\sigma_{a^\dagger})(a-\sigma_a)+{1\over
2}],\label{ham6}\eea with the notation
$H=\hbar(\omega_0+{\g_s/2})a^\dagger a.$
Taking into account the condition (\ref{cpur}),
we see that the mean values of the two Hamiltonians $H$ and $H'$
are equal: $<H>=<H'>.$

In general, the dissipative systems cannot be described by pure
states or by Schr\" odinger equations, because the environment
produces transitions in any state basis. Nevertheless, we will
show that this can happen in very limiting cases, corresponding to
certain special states. In order to find in the general
Karrlein-Grabert model the states which remain pure during the
evolution of the system, we consider the equations of motion for
the second order moments of coordinate and momentum. To obtain
these equations we first derive the evolution equation
(\ref{kar3}) with the coefficients (\ref{dcoef}) in coordinate
representation: \bea i\hbar{\partial\rho\over\partial
t}=-{\hbar^2\over 2M}({\partial^2\over\partial x^2}-
{\partial^2\over\partial y^2})\rho+{M\gamma_q\over
2}(x^2-y^2)\rho-{i\hbar\gamma_p\over
2}(x-y)({\partial\over\partial x}-{\partial\over\partial
y})\rho\nonumber\\ +MD_q(x-y)( {\partial\over\partial
x}+{\partial\over\partial y})\rho
-{i\over\hbar}M^2D_p(x-y)^2\rho.~~~~~~~~~ \label{cooreq}\eea The
first two terms on the right-hand side of this equation generate
purely unitary evolution (with a renormalized potential). The
third term is the dissipative term and the forth is the so-called
"anomalous diffusion" term, which generates a second derivative
term in the phase space representation of the evolution equation,
just like the ordinary diffusion term. The last term is the
diffusive term, which is responsible for the process of
decoherence.
Since the considered dynamics is quadratic, we consider a density
matrix solution of Eq. (\ref{cooreq}) of the form \bea
<x|\hat\rho(t)|y>=({1\over 2\pi\sigma_{qq}(t)})^{1\over 2}
~~~~~~~~~~~~~~~~~~~~~~~~~~~~~~\nonumber \\ \times\exp[-{1\over
8\sigma_{qq}(t)}(x+y)^2 +{i\sigma_{pq}(t)\over
2\hbar\sigma_{qq}(t)}(x^2-y^2)-{1\over 2\hbar^2}
(\sigma_{pp}(t)-{\sigma_{pq}^2(t)\over\sigma_{qq}(t)})(x-y)^2],\label{densol}
\eea which is the general form of Gaussian density matrices (with
zero expectation values of coordinate and momentum).
By direct substitution of $\rho$ into Eq. (\ref{cooreq}), we
obtain the following system of equations satisfied by dispersions
of coordinate and momentum: \bea{d\sigma_{qq}(t)\over dt}={2\over
M} \sigma_{pq}(t),\eea \bea{d\sigma_{pp}(t)\over
dt}=-2\g_p\sigma_{pp}(t)-2M\g_q \sigma_{pq}(t)+2M^2D_p,
\label{eqmo2}\eea \bea{d\sigma_{pq}(t)\over
dt}=-M\g_q\sigma_{qq}(t)+{1\over M}\sigma_{pp}(t)
-\g_p\sigma_{pq}(t)+MD_q.\eea Introducing the notation \bea
X(t)=\left(\matrix{m\sqrt{\g_q}\sigma_{qq}(t)\cr
\sigma_{pp}(t)/m\sqrt{\g_q}\cr \sigma_{pq}(t)\cr}\right)\eea
and solving this system of equations with the method used in Refs.
\cite{rev,ss}, we obtain the solution: \bea
X(t)=T(X(0)-X(\infty))+X(\infty),\label{sol2}\eea
where the matrix $T$ is \bea T=-2{e^{-\g_p t}\over \Omega^2}
\left(\matrix{b_{11}&b_{12}&b_{13}\cr b_{21}&b_{22}&b_{23}\cr
b_{31}&b_{32}&b_{33}\cr}\right),\eea with
time-dependent oscillating functions $b_{ij}$ (i,j=1,2,3) given by
$(\Omega^2=4\g_q-\g_p^2)$: \bea b_{11}=({\g_p^2\over 2}-\g_q)\cos
\Omega t-\g_p{\Omega\over 2}\sin \Omega t-\g_q,\eea \bea
b_{12}=\g_q(\cos \Omega t-1),\eea \bea b_{13}=\sqrt{\g_q}(\g_p\cos
\Omega t-\Omega\sin \Omega t-\g_p),\eea \bea b_{21}=\g_q(\cos
\Omega t-1),\eea \bea b_{22}=({\g_p^2\over 2}-\g_q)\cos \Omega
t+\g_p{\Omega\over 2}\sin \Omega t-\g_q,\eea \bea
b_{23}=\sqrt{\g_q}(\g_p\cos \Omega t+\Omega\sin \Omega
t-\g_p),\eea \bea b_{31}=-\sqrt{\g_q}({\g_p\over 2}\cos \Omega
t-{\Omega\over 2}\sin \Omega t-{\g_p\over 2}),\eea \bea
b_{32}=-\sqrt{\g_q}({\g_p\over 2}\cos \Omega t+{\Omega\over 2}\sin
\Omega t-{\g_p\over 2}),\eea \bea b_{33}=-2\g_q\cos \Omega
t+{\g_p^2\over 2}.\eea The asymptotic values of variances and
covariance have the following expressions:
\bea\sigma_{qq}(\infty)={D_p+\g_pD_q\over\g_p\g_q},~~
\sigma_{pp}(\infty)={M^2D_p\over\g_p},~~
\sigma_{pq}(\infty)=0 \eea
$\sigma_{qq}(0),\sigma_{pp}(0),\sigma_{pq}(0)$. Introducing the
expressions (\ref{dcoef}) for the coefficients $D_q$ and $D_p,$ we
obtain the following equilibrium asymptotic values of the
dispersions: \bea \s_{qq}(\infty)=<q^2>,~~\s_{pp}(\infty)=<p^2>,
~~\s_{pq}(\infty)=0.\label{infvar}\eea
If the asymptotic state is a pure state, then
\bea\s_{qq}(\infty)\s_{pp}(\infty)=<q^2><p^2>={\hbar^2\over
4},\label{infcoh}\eea i. e. it is a squeezed state. Introducing
the expressions of $\s_{qq}(t)$ and $\s_{pq}(t)$ given by (\ref{sol2})
in the condition for purity of states (\ref{cond1}), (\ref{dcoef}),
we have shown, after a long, but straightforward calculation, that
this condition is fulfilled, for any time $t,$ only if the initial
values of dispersions are equal to their asymptotic values:
\bea\s_{qq}(0)=\s_{qq}(\infty),~~\s_{pp}(0)=\s_{pp}(\infty),~~
\s_{pq}(0)=\s_{pq}(\infty).\eea Then it follows from (\ref{sol2})
that $X(t)=X(\infty),$ that is the dispersions have constant
values in time, given by (\ref{infvar}). Therefore, the state
which preserves its purity in time is given by the density matrix
(\ref{densol}), i. e. it is a squeezed state, with the
well-determined constant variances $\s_{qq}, \s_{pp}$
(\ref{infvar}). The fluctuation energy has also a constant value
in time \bea E={1\over 2M}<p^2>+{M\omega_0^2\over 2}<q^2>.\eea At
the same time, the total energy of the open system is given by the
mean value of the Hamiltonian (\ref{kar2}): \bea <H_0>
={1\over 2M}\sigma_{pp}(t)+{M\omega_0^2\over 2}\sigma_{qq}(t)
+{1\over 2M}\sigma_p^2(t)+{M\omega_0^2\over 2}\sigma_q^2(t)
\label{toten}\eea and, since the expectation values of coordinate
and momentum decay exponentially in time \cite{grab}, the energy
is dissipated and reaches the minimum value $E.$ In the particular
case of Agarwal model, the purity condition (\ref{cond5}) shows
that the variance of coordinate must also be constant in time:
\bea \sigma_{qq}(t)={\hbar\over 2M\omega_0\coth({\displaystyle
{\hbar\omega_0\over 2k_BT}})}. \label{rel1}\eea Using this
condition, we find from the equations of motion (41) -- (43)
written for the Agarwal model, when we have to take \bea
D_p={\hbar\omega_0\kappa\over M}\coth({\hbar\omega_0\over
2k_BT}),~~D_q=0,~~\g_p=\kappa,~~\g_q=\omega_0^2,\eea that the
dispersions have to satisfy the following equalities: \bea
\sigma_{pp}(t)={\hbar M\omega_0\over 2}\coth({\hbar\omega_0\over
2k_BT}),~~\s_{pp}(t)=M^2\omega_0^2\s_{qq}(t),~~\sigma_{pq}(t)=0.
\label{rel2}\eea The relations (\ref{rel1}), (\ref{rel2}) are
compatible only if $\coth({\hbar\omega_0/2k_BT})=1$, that is only
when the temperature of the thermal bath is $T=0.$ Then finally we
get \bea \s_{qq}={\hbar\over 2M\omega_0},~~\s_{pp}={\hbar
M\omega_0\over 2},~~\s_{pq}=0\eea and, therefore, in the
particular case of Agarwal model, the usual coherent state is the
only state which remains pure for all times, if the temperature is
$T=0.$ In this case the fluctuation energy of the harmonic
oscillator keeps all the time its minimum value
$E_{min}=\hbar\omega_0/2.$  The relation (\ref{toten}) shows that
in this case the total energy is also dissipated and reaches
$E_{min}.$ The same results can be obtained for the model of
Weidlich and Haake, described by the evolution equation
(\ref{kar10}). Indeed, this model is a particular case (cf. Eqs.
(\ref{kar12})) of the Lindblad model considered in the next
Section and from the purity condition (\ref{cpur}) it follows
again that the coherent state is the only state which preserves
its purity during the evolution in time of the system, for a zero
temperature of the thermal bath. The importance of the states
which preserve their purity in time will become evident in Sec. 5,
in the context of discussing the decoherence phenomenon.

\section{Purity of states in the Lindblad model}

We now consider the condition for purity of states in the Lindblad
model for the damped harmonic oscillator, based on quantum
dynamical semigroups \cite{d,s,l1,l2}.
The most general
Markovian evolution equation preserving the positivity,
hermiticity and trace of $\hat\rho$ can be written as:
\bea{d\hat\rho(t)\over dt}=-{i\over\hbar}[\hat H,\hat\rho(t)]+{1\over 2\hbar}
\sum_{j}([\hat V_{j}\hat\rho(t),\hat V_{j}^\dagger ]+[\hat
V_{j},\hat\rho(t)\hat V_{j}^\dagger ]).\label{lineq}\eea
Here $\hat H$ is the Hamiltonian operator of the system and $\hat V_{j},$
$\hat V_{j}^\dagger $ are
operators on the Hilbert space $\cal H$
of the Hamiltonian which model the interaction with the environment.
In the case of an exactly solvable model for the damped
harmonic oscillator we take the two possible operators $\hat V_{1}$ and
$\hat V_{2}$ linear in $\hat p$ and $\hat q$ \cite{rev,l2,ss}
and the harmonic oscillator
Hamiltonian $\hat H$ is chosen of the general form
\bea   \hat H=\hat H_{0}+{\mu \over 2}(\hat q\hat p+\hat p\hat q),
~~~~\hat H_{0}={1\over 2m}
\hat p^2+{m\omega^2\over 2}\hat q^2.  \label{ham}   \eea
With these choices
the master equation (\ref{lineq}) takes the following form \cite{rev,ss}:
\bea   {d\hat\rho \over dt}=-{i\over \hbar}[\hat H_{0},\hat\rho]-
{i\over 2\hbar}(\lambda +\mu)
[\hat q,\hat\rho \hat p+\hat p\hat\rho]+{i\over 2\hbar}(\lambda -\mu)[\hat p,
\hat\rho \hat q+\hat q\hat\rho]  \nonumber\\
  -{D_{pp}\over {\hbar}^2}[\hat q,[\hat q,\hat\rho]]-{D_{qq}\over {\hbar}^2}
[\hat p,[\hat p,\hat\rho]]+{D_{pq}\over {\hbar}^2}([\hat q,[\hat
p,\hat\rho]]+ [\hat p,[\hat q,\hat\rho]]). ~~~~\label{mast}   \eea
The quantum diffusion coefficients $D_{pp},D_{qq},$ $D_{pq}$ and
the dissipation constant $\lambda$ satisfy the following
fundamental constraints \cite{rev,ss}: $  D_{pp}>0, D_{qq}>0$ and
\bea D_{pp}D_{qq}-D_{pq}^2\ge {{\hbar}^2{\lambda}^2\over 4}.
\label{ineq}  \eea The relation (\ref{ineq}) is a necessary
condition that the generalized uncertainty inequality
(\ref{genun}) is fulfilled.
By using the complete positivity property
it was shown in \cite{ss} that the relation \bea{\rm
Tr}(\hat\rho(t)\sum_j\hat V_j^\dagger\hat V_j)=\sum_j{\rm Tr}
(\hat\rho(t)\hat V_j^\dagger){\rm Tr}(\hat\rho(t)\hat V_j)
\label{has1}\eea represents the necessary and sufficient condition
for $\hat\rho(t)$ to be a pure state for all times $t\ge 0.$ This
equality is a generalization of the pure state condition
[42--44] to all Markovian master equations (\ref{lineq}). If
$\hat\rho^2(t)=\hat\rho(t)$ for all $t\ge 0,$ then there exists a
wave function $\psi\in{\cal H}$ which satisfies a nonlinear
Schr\"odinger equation
with the non-Hermitian Hamiltonian
\bea \hat H'=\hat H+i\sum_j<\psi(t),\hat V_j^\dagger\psi(t)>\hat V_j-{i\over 2}
<\psi(t),\sum_j\hat V_j^\dagger\hat V_j\psi(t)>-{i\over 2}\sum_j
\hat V_j^\dagger\hat V_j. \label{ham1} \eea
For the damped harmonic oscillator
the pure state condition (\ref{has1}) takes the form \cite{ss}
\bea
D_{pp}\sigma_{qq}(t)+D_{qq}\sigma_{pp}(t)-2D_{pq}\sigma_{pq}(t)=
{\hbar^2\lambda\over 2}\label{has2}\eea
and the Hamiltonian (\ref{ham1}) becomes
\bea \hat H'=\hat H+\lambda(\sigma_p(t)\hat q-
\sigma_q(t)\hat p)-{i\over\hbar}
[D_{pp}(\hat q-\sigma_q(t))^2
+D_{qq}(\hat p-\sigma_p(t))^2
\nonumber \\
-D_{pq}((\hat p-
\sigma_p(t))(\hat q-
\sigma_q(t))+(\hat q-
\sigma_q(t))(\hat p-
\sigma_p(t)))-{\lambda\hbar^2\over 2}].~~~~~~~\label{ham2}\eea
It is interesting to remark that the mean value of this Hamiltonian in the state
$\hat\rho(t)$ is equal to the mean value of the Hamiltonian $\hat H$. From
a physical point of view this result is quite natural, since the average value
of the new Hamiltonian $\hat H'$ describing the open system must give the energy
of the open system.


In order to find the Gaussian states which remain pure during the
evolution of the system for all times $t,$ we consider the pure
state condition (\ref{has2}) and the generalized uncertainty
relation
for pure states: \bea
\sigma_{pp}(t)\sigma_{qq}(t)-\sigma_{pq}^2(t) ={\hbar^2\over 4}.
\label{pur}\eea By eliminating $\sigma_{pp}$ between the
equalities (\ref{has2}) and (\ref{pur}), like in \cite{deval}, we
obtain: \bea(\sigma_{qq}(t)-{D_{pq}\sigma_{pq}(t)+{1\over
4}\hbar^2\lambda\over D_{pp}})^2+{D_{pp}D_{qq}-D_{pq}^2\over
D_{pp}^2}[(\sigma_{pq}(t)-{{1\over 4} \hbar^2\lambda D_{pq}\over
D_{pp}D_{qq}-D_{pq}^2})^2 \nonumber\\ +{1\over
4}\hbar^2{D_{pp}D_{qq}-D_{pq}^2-{1\over 4}\hbar^2\lambda^2\over
(D_{pp}D_{qq}-D_{pq}^2)^2}D_{pp}D_{qq}]=0.~~~~~~~~~~~~~~~~~~
\label{dek}\eea Since the diffusion and dissipation coefficients
satisfy the inequality (\ref{ineq}), we obtain from Eq.
(\ref{dek}) the following relations which have to be fulfilled at
any moment of time: \bea
D_{pp}D_{qq}-D_{pq}^2={\hbar^2\lambda^2\over 4}, \label{coe1}\eea
\bea
D_{pp}\sigma_{qq}(t)-D_{pq}\sigma_{pq}(t)-{\hbar^2\lambda\over
4}=0, \label{coe2}\eea \bea
\sigma_{pq}(t)(D_{pp}D_{qq}-D_{pq}^2)-{\hbar^2\lambda\over
4}D_{pq}=0. \label{coe3}\eea From relations (\ref{pur}) and
(\ref{coe1}) -- (\ref{coe3}) it follows that the pure states
remain pure for all times only if the variances have the form:
\bea \sigma_{qq}(t)={D_{qq}\over\lambda},~~
\sigma_{pp}(t)={D_{pp}\over\lambda},~~
\sigma_{pq}(t)={D_{pq}\over\lambda}, \label{coe4}\eea i. e. they
do not depend on time. If these relations are fulfilled, then
the equalities (\ref{has2}), (\ref{pur}) and (\ref{coe1}) are
equivalent.
Using the asymptotic values
of variances for an underdamped oscillator (given by Eqs. (3.53)
in \cite{ss}) and the relations (\ref{coe4}), we obtain the
following expressions of the diffusion coefficients which assure
that the initial pure states remain pure for any $t$
($\Omega^2=\omega^2-\mu^2$): \bea D_{qq}={\hbar\lambda\over
2m\Omega},~~ D_{pp}={\hbar\lambda m\omega^2 \over 2\Omega},~~
D_{pq}=-{\hbar\lambda\mu\over 2\Omega}. \label{coepur}\eea
Formulas (\ref{coepur}) are generalized Einstein relations and
represent typical examples of quantum fluctuation-dissipation
relations, connecting the diffusion with both Planck's constant
and damping constant \cite{d2,loui}. With (\ref{coepur}), the
variances (\ref{coe4}) become
\bea \sigma_{qq}={\hbar\over 2m\Omega},~~
\sigma_{pp}={\hbar m\omega^2 \over 2\Omega},~~
\sigma_{pq}=-{\hbar\mu\over 2\Omega}. \label{varpur}\eea
Then the corresponding state described by a Gaussian Wigner
function is a pure quantum state, namely a correlated coherent
state \cite{dodkur} (squeezed coherent state) with the correlation
coefficient (\ref{corcoe}) $r=-\mu/\omega.$ Given $\sigma_{qq},$
$\sigma_{pp}$ and $\sigma_{pq},$ there exists one and only one
such a state minimizing the uncertainty $\sigma$ (\ref{genun})
\cite{sud}. A particular case of Lindblad model (corresponding to
$\lambda=\mu$ and $D_{pq}=0)$ was considered by Halliwell and
Zoupas by using the quantum state diffusion method \cite{halzou}.
We have considered general coefficients $\lambda$ and $\mu$ and in
this respect our expressions for the diffusion coefficients and
variances generalize also the ones obtained by Dekker and
Valsakumar \cite{deval} and Dodonov and Man'ko \cite{dodman}, who
used models where $\lambda=\mu$ was chosen. If $\mu=0,$ we get
$D_{pq}=0$ from (\ref{coepur}). This case, which was considered in
\cite{ass}, where we obtained a density operator describing a pure
state for any $t,$ is also a particular case of our present
results.
For $\mu=0$, the expressions (\ref{varpur}) become \bea
\sigma_{qq}={\hbar\over 2m\omega},~~ \sigma_{pp}={\hbar
m\omega\over 2},~~ \sigma_{pq}=0,   \label{grovar2}\eea which are
the variances of the ground state of the harmonic oscillator and
the correlation coefficient is $r=0,$ corresponding to the
usual coherent state.

The fluctuation energy of the open harmonic oscillator is \bea
E(t)={1\over 2m}\sigma_{pp}(t)+{1\over 2}m\omega^2\sigma_{qq}(t)+
\mu\sigma_{pq}(t). \label{ener}\eea If the state remains pure in
time, then the variances are given by (\ref{coe4}) and the
fluctuation energy is also constant in time: \bea
E={1\over\lambda}({1\over 2m}D_{pp}+{1\over 2}m\omega^2D_{qq}+\mu
D_{pq}). \label{conen}\eea Minimizing this expression with the
condition (\ref{coe1}), we obtain just the diffusion coefficients
(\ref{coepur}) and $E_{min}=\hbar\Omega/2.$ Therefore, the
conservation of purity of state implies that the fluctuation
energy of the system has all the time the minimum possible value
$E_{min}.$ The total energy of the open system is given by the
mean value of Hamiltonian (\ref{ham}): \bea
<\hat H>={1\over 2m}<\hat p^2> +{m\omega^2\over 2}<\hat q^2>+
{\mu\over 2}<\hat q\hat p+\hat p\hat q>\nonumber \\={1\over
2m}\sigma_{pp}(t)+{1\over 2}m\omega^2\sigma_{qq}(t)+
\mu\sigma_{pq}(t)+{1\over 2m}\sigma_p^2(t)+{m\omega^2\over
2}m\omega^2\sigma_q^2(t)+ \mu\sigma_p(t)\sigma_q(t)\eea and, since
the expectation values of coordinate and momentum decay
exponentially in time \cite{rev,ss}, the energy is dissipated and
reaches its minimum value $E_{min}.$

If the asymptotic state is a Gibbs state \cite{rev,ss},
then the condition (\ref{coe1}) on the diffusion coefficients is
satisfied only if $\mu=0$ and the temperature of the thermal bath
is $T=0.$ Like in the Agarwal and Weidlich-Haake models, discussed
in the previous Section, in this limiting case the influence on
the oscillator is minimal and $E_{min}=\hbar\omega/2,$ which is
the oscillator ground state energy, the correlation coefficient
(\ref{corcoe}) vanishes and therefore the correlated coherent
state (squeezed coherent state) becomes the usual coherent
(ground) state.

The Lindblad equation
with the diffusion coefficients (\ref{coepur}) can
be used only in the underdamped case, when $\omega>\mu.$
Indeed, for the coefficients (\ref{coepur}) the fundamental constraint
(\ref{ineq}) implies that $m^2(\omega^2-\mu^2)D_{qq}^2\ge\hbar^2\lambda^2/4,$
which is satisfied only if $\omega>\mu$.
It can be shown \cite{dodman} that
there exist diffusion coefficients which satisfy the condition
(\ref{coe1}) and make sense for $\omega<\mu,$ but in this overdamped case
we have always $\sigma>\hbar^2/4$ and the state of the oscillator cannot
be pure for any diffusion coefficients.

If we choose the coefficients of the form (\ref{coepur}), then
the equation for
the density operator can be represented in the form (\ref{lineq}) with only
one operator $\hat V,$ which up to a phase factor can be written in the form:
\bea \hat V=\sqrt{{2\over\hbar D_{qq}}}[({\lambda\hbar\over 2}-iD_{pq})\hat q+
iD_{qq}\hat p)],~~~~[\hat V,\hat V^\dagger]=2\hbar\lambda. \label{oneop}\eea

The correlated coherent states (\ref{eigf}) with nonvanishing
momentum average can also be written in the form: \bea
\Psi(x)=({1\over 2\pi\sigma_{qq}})^{1\over 4}\exp[-{1\over
4\sigma_{qq}} (1-{2i\over\hbar}\sigma_{pq})(x-\sigma_q)^2+{i\over
\hbar}\sigma_px] \label{ccs}\eea and the most general form of
Gaussian density matrices compatible with the generalized
uncertainty relation (\ref{genun}) is the following: \bea
<x|\hat\rho|y>=({1\over 2\pi\sigma_{qq}})^{1\over 2} \exp[-{1\over
2\sigma_{qq}}({x+y\over 2}-\s_q(t))^2\nonumber \\
+{i\sigma_{pq}\over\hbar\sigma_{qq}}({x+y\over
2}-\s_q)(x-y)-{1\over 2\hbar^2}
(\sigma_{pp}-{\sigma_{pq}^2\over\sigma_{qq}})(x-y)^2+{i\over\hbar}\s_p(x-y)]\label{ccd}.
\eea
These matrices correspond to the correlated coherent states
(\ref{ccs}) if $\sigma_{qq}, \sigma_{pp}$ and $\sigma_{pq}$ in
(\ref{ccd}) satisfy the equality (\ref{genun}), in particular if
the variances are taken of the form (\ref{varpur}).
Consider now the harmonic oscillator initially in a correlated coherent state
of the form (\ref{ccs}), with the corresponding Wigner
function (\ref{corwig}).
For an environment described by the diffusion coefficients (\ref{coepur}),
the Wigner function at time $t$ is given by
\bea   W(p,q,t)={1\over \pi\hbar}
~~~~~~~~~~~~~~~~~~~~~~~~~~~~~~~~~ \nonumber \\
\times\exp\{-{2\over \hbar^2}[\sigma_{pp}(q-\sigma_q(t))^2+
\sigma_{qq}(p-\sigma_p(t))^2-2\sigma_{pq}(q-\sigma_q(t))(p-\sigma_p(t))]\},
\label{wig} \eea
with the constant variances (\ref{varpur}).
The correlated coherent state (squeezed coherent state) remains a
correlated coherent state with variances constant in time and with
$\sigma_q(t)$ and $\sigma_p(t)$ giving the average time-dependent
location of the system along its trajectory in phase space. In the
long-time limit $\sigma_q(t)=0,$ $\sigma_p(t)=0$ and then we have
\bea
<x|\hat\rho(\infty)|y>= ({m\Omega\over\pi \hbar})^{1\over 2}
\exp\{-{m\over 2\hbar}[\Omega(x^2+y^2)+i\mu(x^2-y^2)]\}.
\label{coorinf}\eea The corresponding Wigner function has the form
\bea   W_{\infty}(p,q)={1\over \pi\hbar} \exp[-{2\over
\hbar\Omega}({p^2\over 2m}+{m\over 2}\omega^2q^2+\mu pq)].
\label{wiginf}\eea

\section{Entropy and decoherence}

Besides the von Neumann entropy $S$ (\ref{entro}), (\ref{entro2}),
there is another quantity which can measure the degree of mixing
or purity of quantum states. It is the linear entropy $S_l$
defined as \bea S_l={\rm Tr}(\hat\rho-\hat\rho^2)=1-{\rm
Tr}\hat\rho^2.\label {entro3}\eea For pure states $S_l=0$ and for
a statistical mixture $S_l>0.$ As it is well-known, the increasing
of the linear entropy $S_l$ (as well as of von Neumann entropy
$S$) due to the interaction with the environment is associated
with the decoherence phenomenon (loss of quantum coherence), given
by the diffusion process \cite{paz1,paz2}. Dissipation increases
the entropy and the pure states are converted into mixed states.
The rate of entropy production is given by \bea {\dot
S}_l(t)=-2{\rm Tr}(\hat\rho\dot{\hat \rho})=-2{\rm Tr}
(\hat\rho{\cal L}(\hat\rho)), \label{entpro1}\eea where $\cal L$
is the evolution operator. According to Zurek's theory
\cite{paz1,paz2}, the maximally predictive states are the pure
states which minimize the entropy production in time.
These states remain least affected by the openness of the system
and form a "preferred set of
states" in the Hilbert space of the system, known as the
"pointer basis". Decoherence is the mechanism which selects these
preferred states -- the most stable ones under the evolution
in the presence of the environment.

For the models of the damped harmonic oscillator considered in
this paper, we can obtain the expressions for the rate of entropy
production given by Eq. (\ref{entpro1}). For Gaussian states the
linear entropy (\ref{entro3}) becomes \bea S_l(t)=1- {1\over
\nu},~~~ \nu={2\over\hbar}\sqrt{\s}\label{gausent}\eea and then
the time derivative of the linear entropy is given by
 \bea
\dot S_l(t)={1\over \nu^2}{d\nu\over dt}={\hbar\over
4\s\sqrt{\s}}[{d\s_{qq}(t)\over dt} \s_{pp}(t)+{d\s_{pp}(t)\over
dt} \s_{qq}(t)-2{\s_{pq}(t)\over dt}\s_{pq}(t)].\eea From the
system of equations (41) -- (43) for the Karrlein-Grabert model we
obtain \bea \dot S_l(t)={\hbar\over
2\s\sqrt{\s}}[M^2D_p(t)\s_{qq}(t)-MD_q(t)\s_{pq}(t)-\g_p(t)\s].
\label{kgrate}\eea Suppose at the initial moment of time $t=0$ the
state is pure.
When the conditions (\ref{cond1}), (\ref{cond2}) for purity of
states are fulfilled for all $t$, the expression of the rate of
linear entropy becomes
 \bea
{\dot S_l(t)}={4\over
\hbar^2}[M^2D_p(t)\s_{qq}(t)-MD_q(t)\s_{pq}(t)-{\hbar^2\over
4}\g_p(t)]=0 \label{rate1}\eea and then the entropy production has
its minimum value $S_l=0.$ For the thermal initial condition with
the coefficients (\ref{therm}), the rate of entropy production is
given by \bea \dot S_l(t)={\hbar\over
2\s\sqrt{\s}}[\g_p(t)<p^2>\sigma_{qq}(t)-(M\g_q(t)<q^2>-
{<p^2>\over M}) \sigma_{pq}(t)-\g_p(t)\s], \label{rate2}\eea  for
strictly Ohmic damping it is \bea \dot S_l(t)={\hbar\over 2
\s\sqrt{\s}}[\gamma<p^2>\sigma_{qq}(t)-
(M\omega_0^2<q^2>-{<p^2>\over M}) \sigma_{pq}(t)-\g\s]
\label{rate3}\eea and for Drude damping the rate of entropy
production is also given by an expression like (\ref{rate2}),
where now $\gamma_p=2\alpha$ and $\gamma_q=\alpha^2+\eta^2.$ When
the condition for purity is fulfilled for any $t,$ the values of
the rate of linear entropy given by (\ref{rate2}), (\ref{rate3})
become also 0. According to the results of Sec. 3, if the
condition for purity of states is fulfilled for any $t$ in the
Karrlein-Grabert model, then the Gaussian state will be a pure
squeezed state, with constant in time variances.
At the same time the rate of linear entropy production vanishes
and, therefore, according to the Zurek's theory of decoherence,
the most stable states are the pure squeezed states, with constant
variances. The same conclusion is valid for the weak damping
model, given by the master equation (\ref{kar8}), for which the
rate of entropy production has the expression \bea \dot
S_l(t)={\hbar^2\over 2\s\sqrt{\s}}[K_c\sigma_{qq}(t)+{K_s\over
M\omega_0} \sigma_{pq}(t)-\g_c{\s\over\hbar}],\label{rate4}\eea
while for the Agarwal model given by the master equation
(\ref{kar9}) we obtain \bea \dot S_l(t)={\hbar^2\kappa\over 2
\s\sqrt{\s}}[M\omega_0\coth({\hbar\omega_0\over 2k_BT})
\sigma_{qq}(t)-{2\s\over\hbar}]. \label{rate5}\eea Analogously,
for Eq. (\ref{kar10}) of Weidlich and Haake, the rate of entropy
production is given by \bea \dot S_l(t)={\hbar^2\g_c\over
8\s\sqrt{\s}}[(M\omega_0\sigma_{qq}(t)+{\sigma_{pp}(t) \over
M\omega_0})\coth({\hbar\omega_0\over 2k_BT})-{4\s\over\hbar}]
\label{rate6} \eea and, according to the results of Sec. 3, for
Agarwal and Weidlich-Haake models, the usual coherent states are
the most stable ones under evolution in the presence of the
environment. Using Eq. (\ref{entpro1}) for the Lindblad equation
(\ref{mast}), we obtain the following rate of entropy production:
\bea \dot S_l(t)={4\over\hbar^2} [D_{pp}{\rm Tr}(\hat\rho^2\hat
q^2-\hat\rho\hat q\hat\rho\hat q) ~~~~~~~~~~~~~~~~~~~~\nonumber \\
 +D_{qq}{\rm Tr}(\hat\rho^2\hat p^2-\hat\rho\hat p\hat\rho\hat p) -
D_{pq}{\rm Tr}(\hat\rho^2(\hat q\hat p+\hat p\hat q)-
2\hat\rho\hat q\hat\rho\hat p) - {\hbar^2\lambda\over 2}{\rm
Tr}(\hat\rho^2)] \label{entpro3} \eea or, using Eq.
(\ref{gausent}) for Gaussian states, \bea \dot S_l(t)={\hbar\over
2\s\sqrt{\s}}[D_{pp}\s_{qq}(t)+D_{qq}\s_{pp}(t)-2D_{pq}\s_{pq}(t)
-2\lambda\s].\label{ratent}\eea If the initial state is pure, then
according to the complete positivity property of the Lindblad
model we have \bea \dot
S_l(0)={4\over\hbar^2}[D_{pp}\sigma_{qq}(0)+D_{qq}\sigma_{pp}(0)-
2D_{pq}\sigma_{pq}(0)-{\hbar^2\lambda\over 2}]\ge 0,\eea which
means that the linear entropy can only increase, so that the
initial pure state becomes mixed. When the state remains pure, Eq.
(\ref{ratent}) becomes, cf. Eq. (\ref{has2})
: \bea \dot
S_l(t)={4\over\hbar^2}[D_{pp}\sigma_{qq}(t)+D_{qq}\sigma_{pp}(t)-
2D_{pq}\sigma_{pq}(t)-{\hbar^2\lambda\over 2}]=0\label
{entpro4}\eea
and, therefore, the entropy production will be $S_l=0$. Since the
only initial states which remain pure for any $t$ are the
correlated coherent states, we can state that in the Lindblad
theory these states are the maximally predictive states. The
present results, obtained in the framework of Karrlein-Grabert and
Lindblad models, generalize the previous results which assert that
for many models of quantum Brownian motion in the high temperature
limit the usual coherent states correspond to minimal entropy
production and, therefore, they are the maximally predictive
states. As we have seen, such coherent states can be obtained in
the Lindblad model as a particular case of the correlated coherent
states by taking $\mu=0,$ so that the correlation coefficient
(\ref{corcoe}) $r=0.$ Namely, Paz, Habib and Zurek
\cite{paz1,paz2} considered the harmonic oscillator undergoing
quantum Brownian motion in the Caldeira-Leggett model and
concluded that the minimizing states which are the initial states
generating the least amount of von Neumann or linear entropy and,
therefore, the most predictable or stable ones under evolution in
the presence of an environment, are the ordinary coherent states.
Using an information-theoretic measure of uncertainty for quantum
systems, Anderson and Halliwell showed in \cite{AndH} that the
minimizing states are certain general Gaussian states.
Anastopoulos and Halliwell \cite{AnH} offered an alternative
characterization of these states by noting that they minimize the
generalized uncertainty relation. According to this assertion, we
can say that in the Lindblad model the correlated coherent states
are the most stable ones which minimize the generalized
uncertainty relation (\ref{genun}). Our result confirms that one
of \cite{AnH}, where the model for the open quantum system
consists of a particle moving in a harmonic oscillator potential
and linearly coupled to an environment consisting of a bath of
harmonic oscillators in a thermal state. We remind that the
Caldeira-Leggett model considered in \cite{paz1,paz2} violates the
positivity of the density operator at short time scales
\cite{Amb,dio}, whereas in the Lindblad model the property of
positivity is always fulfilled.

The rate of predictability loss, measured by the rate of linear
entropy increase, is also calculated in the framework of Lindblad
theory for the damped harmonic oscillator by Paraoanu and Scutaru
\cite{par}, who have shown that, in general, the pure or mixed
state which produces the minimum rate of increase in the area
occupied by the system in the phase space is a quasi-free state
which has the same symmetry as that induced by the diffusion
coefficients. For isotropic phase space diffusion, coherent states
(or mixture of coherent states) are selected as the most stable
ones. In order to generalize the results of Zurek and
collaborators, the entropy production was also considered by
Gallis \cite{gal} within the Lindblad theory of open quantum
systems, treating environment effects perturbatively. Gallis
considered the particular case with $D_{pq}=0$ and found out that
the squeezed states emerge as the most stable states for
intermediate times compared to the dynamical time scales. The
amount of squeezing decreases with time, so that the coherent
states are most stable for large time scales. For $D_{pq}\not=0$
our results generalize the result of Gallis and establish that the
correlated coherent states are the most stable ones under the
evolution in the presence of the environment.

\section{Summary}

In the present paper we have first considered the generalized
quantum master equations derived by Karrlein and Grabert
\cite{karr} for the microscopic model of a harmonic oscillator
coupled to a harmonic bath. We have obtained the conditions for
purity of states for different initial conditions and different
types of damping, including strictly Ohmic, Drude and weak
coupling cases, Agarwal and Weidlich-Haake models. We have shown
that the states which remain pure all the time are the pure
squeezed states with well-determined constant in time variances.
For pure states, we have also derived the corresponding
generalized Schr\" odinger-type nonlinear equations. Then we have
studied the one-dimensional harmonic oscillator with dissipation
within the framework of Lindblad theory and have shown that the
only states which stay pure during the evolution of the system are
the correlated coherent states, under the condition of a special
choice of the environment coefficients, so that the variances and
covariance are constant in time. We have also obtained the
expressions for the rate of entropy production in the considered
models and have shown that the states which preserve their purity
in time are also the states which minimize the entropy production
and, therefore, they are connected with the decoherence
phenomenon. According to the Zurek's theory of decoherence, in
Karrlein-Grabert and Lindblad models, as well as in the considered
particular models, these states are the most stable ones under the
evolution of the system in the presence of the environment. In a
next work in the framework of these theories we plan to discuss in
more details the connection between uncertainty, decoherence and
correlations of open quantum systems with their environment.

{\bf Acknowledgements}

Two of us (A. I. and A. S.) are pleased to express their sincere gratitude
for the hospitality at the Institut f\"ur Theoretische
Physik in Giessen and to Prof. H. Scutaru for enlightening discussions and
reading the manuscript. A. I. and A. S. also gratefully acknowledge
financial support
by the Deutsche Forschungsgemeinschaft (Germany).
The authors would also like to thank the referee for his comments and
recommendations.


\end{document}